\documentclass[aps,prc,twocolumn,showpacs,superscriptaddress,footinbib]{revtex4-1}

\usepackage{graphicx}
\usepackage{calrsfs}
\usepackage{dcolumn}
\usepackage{bm}
\usepackage{enumerate}
\usepackage{amsmath}

\usepackage[normalem]{ulem}  
\usepackage[colorlinks]{hyperref}

\begin{document}


\title{Size properties of the largest fragments produced in the framework of the statistical multifragmentation model}

\author{S.R. Souza}
\affiliation{Instituto de F\'\i sica, Universidade Federal do Rio de Janeiro Cidade Universit\'aria, 
\\Caixa Postal 68528, 21941-972 Rio de Janeiro-RJ, Brazil}
\affiliation{Departamento de F\'\i sica, ICEx, Universidade Federal de Minas Gerais,
\\Av.\ Ant\^onio Carlos, 6627, 31270-901 Belo Horizonte-MG, Brazil}
\author{R. Donangelo}
\affiliation{Instituto de F\'\i sica, Universidade Federal do Rio de Janeiro Cidade Universit\'aria, 
\\Caixa Postal 68528, 21941-972 Rio de Janeiro-RJ, Brazil}
\affiliation{Instituto de F\'\i sica, Facultad de Ingenier\'\i a, Universidad de la Rep\'ublica, 
Julio Herrera y Reissig 565, 11.300 Montevideo, Uruguay}

\date{\today}

\begin{abstract}
We study the size properties of the largest intermediate mass 
fragments in each partition mode, produced in the prompt statistical 
breakup of a thermally equilibrated nuclear source, at different 
temperatures.
We find that an appreciable amount of events have primary intermediate 
mass fragments of similar sizes.
Our results suggest that, depending on the temperature of the fragmenting 
source, their production may be much larger than what would be 
expected from considerations based on purely combinatorial 
arrangements of the nucleons in the fragmenting system.
We also find that the isospin composition of the largest fragments is 
sensitive to their rank size within the event.
We suggest that experimental analyses, conceived to reconstruct the 
breakup configuration, should be employed to investigate the validity of 
our findings.

\end{abstract}

\maketitle

\begin{section}{Introduction}
\label{sect:introduction}
The understanding of the dynamics of the violent collision
of two heavy ions at bombarding energies ranging from a few tens to 
a few hundreds MeV per nucleon, leading to many 
nuclear fragments in the exit channel, has 
been a challenge for both theorists and experimentalists 
during the last few decades \cite{Moretto1993,reviewSubal2001,BorderiePhaseTransition2008,
PhaseTransitionBorderie2019,reviewBaoAnLi2008}.
Although a copious production of complex fragments in central and
mid-central collisions has been clearly established, the mechanisms 
leading to it have been interpreted in different scenarios.
Indeed, many properties of the fragments observed experimentally have 
been explained by statistical models \cite{Moretto1993,Bondorf1995,GrossPhysRep,reviewSubal2001,
BettyPhysRep2005}, 
whereas many features are also adequately described by dynamical 
treatments, which range from classical and semi-classical formulations 
\cite{AichelinPhysRep,BonaseraBUU,reviewBaoAnLi2008} 
to quantum approaches \cite{fmd1997,AMDReview2004}.
Hybrid treatments have also been employed to match different 
different stages of the reaction, so that one approach provides 
the input information to the other 
\cite{BorderiePhaseTransition2008,BotvinaMishustin2006,CoolDyn2006} 
or act concomitantly \cite{twoPcor2019}, merging different mechanisms.

Among the different predictions made by the dynamical models, 
the emergence of regions of negative compressibility of the 
nuclear matter, {\it i.e.} regions of spinodal instability, 
allows the development of instabilities that would lead to 
the breakup of the system 
\cite{BurgioSpinodalInstabilities1994,spinodalBaldo95,
spinodalInstabilitiesChomaz,spinodalColonna97}.
A salient feature of this mechanism is the formation of nearly 
sized fragments \cite{BurgioSpinodalInstabilities1994,spinodalBaldo95,
spinodalInstabilitiesChomaz,spinodalColonna97,FragsSizeBorderie2018,
BorderiePhaseTransition2008,PhaseTransitionBorderie2019}.
The detection of such events has challenged experimentalists since 
different factors make it very difficult to draw precise conclusions 
on this respect.
For instance, the deexcitation of these fragments after the breakup 
could obscure their resemblance when they were formed.
Experimental indications of the existence of such events have been 
reported recently \cite{FragsSizeBorderie2018} and the discussion 
on whether they are due to merely combinatorial 
arrangements of a finite number of nucleons, or are actually due 
to the development of spinodal instabilities, has been addressed 
based on correlations proposed in Ref.\ \cite{fragSizeCorrelations2002}.

In this work, we examine the properties of the largest fragments formed 
in the prompt statistical breakup of a nuclear source in thermal 
equilibrium. 
We investigate whether the production of similar sized fragments 
is dominated by combinatorial arrangements of nucleons 
or by the statistical weights associated with the accessible phase
space.
In order to elimitate difficulties associated with the incomplete 
sampling of the huge partition space, we use a version of the 
Statistical Multifragmentation Model (SMM) \cite{smm1,smm2,smm4} 
based on the exact recurrence formulae developed in 
Refs.\ \cite{ChaseMekjian1995,SubalMekjian}.
Focusing on partitions with only a few large fragments, 
the equations derived in this work allow the calculation 
of the individual properties of each of these partitions. 
The manuscript is organized as follows.
The main features of the SMM are recalled in Sect.\ \ref{sect:model}, 
where the formulae used in this work are derived.
The results are presented in Sect.\ \ref{sect:results} and the main 
conclusions are drawn in Sect.\ \ref{sect:conclusions}.

\end{section}
 
\begin{section}{Theoretical framework}
\label{sect:model}
The SMM is described in detail in the original works 
where it has been formulated \cite{smm1,smm2,smm4}.
Modifications to include improved binding energies and internal 
Helmholtz free energies are also carefully discussed in 
refs.\ \cite{ISMMmass,ISMMlong}.
Therefore, in subsect. \ref{subsection:SMM}, we briefly sketch 
the main points useful in the discussion below and focus, 
in subsect. \ref{subsect:Subal}, on the derivation of the 
formulae employed in our analysis.

It is assumed that a thermal equilibrated source of mass and atomic 
numbers $A_0$ and $Z_0$, respectively, is formed at temperature $T$ 
and density $\rho$ and that it undergoes a prompt statistical breakup.
As in previous studies \cite{ISMMlong}, we adopt $\rho=\rho_0/6$, 
where $\rho_0$ corresponds to the normal nuclear matter density.
In order to examine the sensitivity of the results to the excitation 
of the system, different values of the breakup temperature are used 
in Sect.\ \ref{sect:results}.

\begin{subsection}{The SMM}
\label{subsection:SMM}
Partitions are generated according to mass and charge conservation, 
so that the multiplicities $\{n_i\}$ of fragments of mass and atomic numbers $a_i$ and $z_i$, respectively, are subject to the constraints:

\begin{equation}
\sum_i n_i a_i=A_0\quad{\rm and}\quad \sum_i n_i z_i=Z_0\;.
\label{eq:conservation}
\end{equation}

In the canonical formulation of the model \cite{smmIsobaric}, the 
statistical weight associated with a fragmentation mode 
$f=\{(a_1,z_1)\cdots(a_m,z_m)\}$, $m\equiv\sum_i n_i$, fulfilling 
the above constraints, is given by the partition function:

\begin{equation}
\Omega_f=\exp\left[-\frac{F_f(T,V)}{T}\right]\;,
\label{eq:partFunc}
\end{equation}

\noindent
where $V$ corresponds to the breakup volume and $F_f(T,V)$ symbolizes 
the Helmholtz free energy associated with the fragmentation mode 
\cite{ISMMlong,smmIsobaric}.

As discussed in Ref.\ \cite{smm4}, the number of different partitions 
rapidly becomes prohibitively large to allow the direct generation 
of all of them.
For this reason, the standard SMM adopts a Monte Carlo strategy, 
in which different fragmentation modes are generated based on 
the combinatorial weight $W_f^{-1}$ of a partition $f$ 
\cite{smm4}.
In this way, the average value of an observable $O$, associated 
with the primary hot fragments, is given by:

\begin{equation}
\langle O\rangle =\frac{\sum_f O_f\,\Omega_f\, W_f}{\sum_f \Omega_f\,W_f}\;.
\label{eq:observable}
\end{equation}

Since most of the primary fragments are very excited, their yields 
will be significantly affected in most cases 
\cite{ISMMlong,smmde2013,EnergySpectra2018,twoPcor2019}.
However, as we are interested in the system's properties at the point 
it disassembles, we will not consider their deexcitation.

It should be stressed that the combinatorial factor which appears 
in the above equation is meant to correct for the fact that the 
partitions $\{f\}$, which enter into Eq.\ (\ref{eq:observable}), 
are not generated by the Monte Carlo sampling with equal probability.
Rather, they are selected according to a distribution 
$W_f^{-1}$ \cite{smm4}.
Therefore, the average value of an observable, calculated considering 
only the possible combinatorial arrangements of $A_0-Z_0$ neutrons and $Z_0$ protons, assuming that they occur with equal probability, 
{\it i.e.} disregarding all other physical effects, is given by:

\begin{equation}
\langle \tilde O\rangle =\frac{\sum_f O_f W_f}{\sum_f W_f}\;.
\label{eq:observableMC}
\end{equation}

In order to evaluate the enhancement or suppression of 
$\langle O\rangle$ with respect to what would be obtained 
considering only constraints due to this combinatorial 
arrangement, one may calculate the ratio:

\begin{equation}
R_O = \frac{\langle O\rangle}{\langle \tilde O\rangle}\;.
\label{eq:ratioMC}
\end{equation}

\noindent
However, owing to the huge number of partitions in the case of 
systems of actual interest, average values of observables 
may be subject to large fluctuations, if their main contributions 
arise from rare events.
Thus, ratios based on such observables may be significantly affected 
by the rather reduced sampling of the configurations.
This assertion remains valid even if a very large (but practically 
feasible) number of partitions is generated, as they would consider 
only a very small fraction of the total set.
This is particularly important if the denominator of the ratio is small.
We have checked that this indeed happens in the case of the observables 
discussed in the next section, even if as many as $10^9$ 
Monte Carlo partitions are sampled.

\end{subsection}

\begin{subsection}{Recurrence relations}
\label{subsect:Subal}
To eliminate this difficuty, we resort to the formulation developed 
by Das Gupta and Mekjian \cite{ChaseMekjian1995,SubalMekjian}, in 
which different observables may be exactly calculated through 
recurrence relations.
More specifically, the statistical weight associated with a source 
$(A_0,Z_0)$ is written as:

\begin{equation}
\Omega_{A_0,Z_0}=\sum_{f\in F_0}\prod_{i\in f}\frac{\omega_i^{n_i}}{n_i!}\;,
\label{eq:Omega}
\end{equation}

\noindent
where $F_0$ symbolizes the set of partitions consistent with the 
constraints expressed by Eq.\ (\ref{eq:conservation}) and

\begin{equation}
\omega_i=\left(\frac{g_i V_f}{\lambda_T^3}A_i^{3/2}\right)
\exp\left(-{\cal F}_i/T\right)\;.
\label{eq:omega}
\end{equation}

\noindent
In the above equation, $g_i$ denotes the spin degeneracy factor of 
the species $(a_i,z_i)$, $\lambda_T=\sqrt{2\pi\hbar^2/mT}$, $m$ is 
the nucleon mass and ${\cal F}_i$ symbolizes the contribution of 
species $i$ to the total Helmholtz free energy.
It contains terms associated with its binding energy and to the 
Wigner-Seitz corrections to the Coulomb energy \cite{smm1}, besides 
others associated with its internal excitation \cite{isoMassFormula2008}.

Das Gupta and Mekjian \cite{ChaseMekjian1995,SubalMekjian} realized 
that very efficient recursion relations could be obtained from 
Eq.\ (\ref{eq:Omega}) and derived the following result:

\begin{equation}
\Omega_{A_0,Z_0}=\sum_{(a_i,z_i) \in S_0}\frac{a_i}{A_0} \omega_i\, 
\Omega_{A_0-a_i,Z_0-z_i}\;,
\label{eq:OmegaRec}
\end{equation}

\noindent
where $S_0$ corresponds to the set composed of all species 
$\{(a_i,z_i)\}$ for which $(a_i,z_i) \le (A_0,Z_0)$.

In the same vein, we extend this idea to calculate the probability 
of observing a particular partition $f\in F_0$, which contains a 
subset $s = \{n_1,\cdots,n_{m_0}\}$ of $M=\sum_{i=1}^{m_0}n_i$ 
fragments, $s\subset f$, which fulfills a condition ${\cal C}$:

\begin{equation}
\tilde P_{s,f}=\frac{1}{\Omega_{A_0,Z_0}}
\left(\prod_{i\in s}\frac{\omega_i^{n_i}}{n_i!}\right)
\left(\prod_{k\in f\setminus s}\frac{\omega_k^{n_k}}{n_k!}\right)\;.
\label{eq:pstilde}
\end{equation}

\noindent
We denote the subset made up of these particular partitions $f$ 
by $F_{0,s}$.
By defining

\begin{equation}
\Omega^*_s\equiv \prod_{i\in s}\frac{\omega_i^{n_i}}{n_i!}\;,
\label{eq:omegas}
\end{equation}

\begin{equation}
\tilde \Omega_{A_0,Z_0}^s\equiv \sum_{f\in F_{0,s}}\,
\prod_{k\in f\setminus s}\frac{\omega_k^{n_k}}{n_k!}\;,
\label{eq:omegas}
\end{equation}

\noindent
the probability of observing the set of fragments $s$ among all 
possible partitions is given by:

\begin{equation}
P_s=\sum_{f\in F_{0,s}}\tilde P_{s,f}=\frac{1}{\Omega_{A_0,Z_0}}\Omega^*_s\,\tilde\Omega_{A_0,Z_0}^s\;.
\label{eq:ps}
\end{equation}

\noindent
If $M$ is small and the fragmenting source is not too large, 
the above equations can be evaluated numerically, considering all the 
possible partitions, if the condition ${\cal C}$ is strict enough.
For instance, considering $A_0=189$, $Z_0=83$ (which in our 
implementation of the SMM gives 3209 species in set $S_0$), 
fragmentation modes containing only $M$ fragments with $z \ge 5$,
the subsets $s$ with $M=3$ 
and $4$ such fragments have $2.43961 \times 10^8$ and 
$1.62527\times 10^{10}$ partitions, respectively.
Although the evaluation of observables associated with the selected 
fragments within these partitions is time consuming, it is a feasible 
task using the present computational resources.
One should note that the partitions which are actually calculated 
individually are the subsets $s$, entering into Eq.\ (\ref{eq:ps}), 
{\it i.e.} $\Omega^*_s$.
The remaining contribution is taken into account by 
$\tilde \Omega_{A_0,Z_0}^s$, which is evaluated recursively through 
Eq.\ (\ref{eq:OmegaRec}), including only species which are not in $s$.
Otherwise, the number of partitions would be too large to allow the 
direct evaluation of the sums which enter into the above expressions, 
even in the rather particular cases exemplified above.

\end{subsection}

\end{section}

\begin{figure}[tbh]
\includegraphics[width=8.5cm,angle=0]{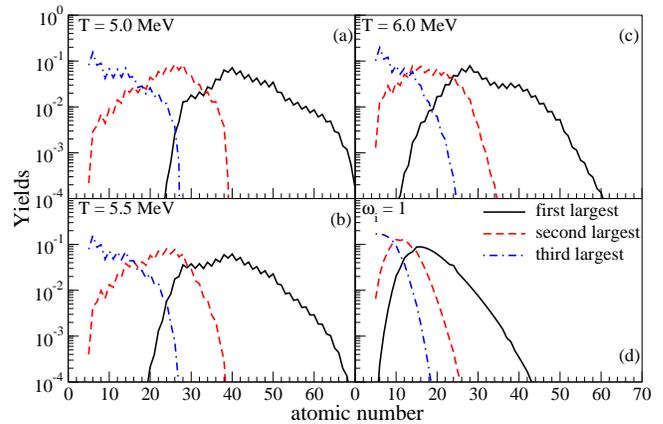}
\caption{\label{fig:zDist3} (Color online) Charge distribution of 
the first, second, and third, largest fragments in events in which 
$M=3$ fragments have atomic number $z \ge 5$, for different breakup 
temperatures, panels (a)-(c). 
In panel (d) the fragments appear in the partitions according 
to rules dictated by the combinatorial arrangements of 
$A_0-Z_0$ neutrons and $Z_0$ protons. For details, see the text.}
\end{figure}

\begin{section}{Results}
\label{sect:results}
We now apply the model to study some properties of the fragments 
produced in the fragmentation of a source of size $A_0=189$ and $Z_0=83$.
As in the calculations reported in Ref.\ \cite{FragsSizeBorderie2018}, 
this corresponds to 80\% of the $^{124}{\rm Xe}+^{112}{\rm Sn}$ system, 
studied experimentally in that work.

We start by considering the charge distribution of the fragments 
observed in partitions $f \in F_{0,s}$ which fulfill the condition 
${\cal C}$ that the $M$ largest fragments have atomic numbers 
$z \ge 5$, whereas the others have smaller atomic numbers.
We calculate the average yields of the $k$-{\it th} largest fragments 
using $P_s$, given by Eq.\ (\ref{eq:ps}):

\begin{equation}
\langle Y_k(z)\rangle =\left(\sum_{\substack{s\in S^*_M\\ z_k\in s}} P_s\, 
\delta_{z,z_k}\right)/\left(\sum_{s\in S_M} P_s\right)\;,
\label{eq:yrank}
\end{equation}

\noindent
where $S^*_M$ denotes the subset of fragments within the partitions 
fulfilling the condition ${\cal C}$ and $\delta$ is the Kronecker delta.

This is displayed in panels (a)-(c) in Fig.\ \ref{fig:zDist3} for $M=3$ 
and breakup temperatures $T= 5.0,\, 5.5,$ and 6.0 MeV.
One sees that, as the temperature rises, the distributions of the 
first, second, and third largest fragments become narrower and their 
peaks shift towards small $z$ values, while the separation between 
them diminishes.
The large separation between the peaks indicates that the fragment 
sizes are appreciably different in most cases.
However, the overlap between the distributions suggests that there 
are partitions in which the 3 fragments have similar atomic 
numbers.
Analogous conclusions hold for $M=4$.
These properties are in qualitative agreement with the experimental 
findings reported in Ref.\ \cite{FragsSizeBorderie2018}, but our 
results cannot be directly compared to those data as we focus on 
the system's configuration at the breakup stage and do not consider 
the subsequent deexcitation of the fragments.

The nonvanishing overlap between the distributions displayed in 
panels (a)-(c) of Fig.\ \ref{fig:zDist3} shows that large fragments 
of similar sizes may also be produced in the statistical breakup 
of the system.
Hence, the existence of events with this property is not an exclusive 
feature of the disassembly by spinodal instabilities, predicted by 
dynamical mean field calculations \cite{BurgioSpinodalInstabilities1994,spinodalBaldo95,
spinodalInstabilitiesChomaz,spinodalColonna97}.
However, there still remains the question of whether the existence 
of such events in statistical multifragmentation merely reflects the 
constraints associated with the mass/charge conservation laws and the 
combinatorial arrangements of the nucleons.
To examine this point, we show in panel (d) the charge distribution 
of the $M$ fragments obtained assuming $\omega_i=1$ in 
Eqs.\ (\ref{eq:pstilde})-(\ref{eq:omegas}).
In this way, the fragments contribute the same weight to the 
partition, except for the factors $(n_i!)^{-1}$ associated with the 
proper counting of identical fragments.
This is also adopted in different methods used to generate partitions 
\cite{smm4,GrossPhysRep}.
One sees that the qualitative features observed in panels (a)-(c), 
considering the full statistical weight, are also present in this 
scenario and that the shape of the distributions is similar in both 
cases.
However, the distributions shown in panel (d) are temperature independent.
They are determined by the system size and the species included in the 
set $S_0$.
The distributions tend to become more and more similar to the full 
statistical ones as the temperature increases and the latter 
distributions shift to lower $z$ values.

As in Ref.\ \cite{FragsSizeBorderie2018}, we now consider the first 
and the second moments of the charge distribution of these $M$ 
fragments in the subset $s$: 

\begin{equation}
\langle z\rangle_s = \frac{1}{M}\sum_{i\in s} z_i
\label{eq:aveZ}
\end{equation}

\noindent
and

\begin{equation}
\sigma_{z,s}=\left[\frac{1}{M}
\sum_{i\in s}(z_i-\langle z\rangle)^2\right]^{1/2}\;.
\label{eq:sigma}
\end{equation}

\noindent
From them, and using Eq.\ (\ref{eq:ps}), we build the frequency 
with which sets of $M$ fragments are observed 
with average value $\langle z\rangle$ and variance $\sigma_z$:

\begin{equation}
Y(\langle z\rangle,\sigma_z) = \left(\sum_{s\in S^*_M} P_s\, 
\Delta_{\langle z\rangle_s,\sigma_{z,s}}\right)/
\left(\sum_{s\in S_M} P_s\right)
\label{eq:Y}
\end{equation}

\noindent
where
 
\begin{equation}
\Delta_{\langle z\rangle_s,\sigma_{z,s}} = 
\begin{cases}
1,\, {\rm if}\quad \delta_{z,s}\le\delta_z/2 \; {\rm and}\; 
\delta_{\sigma_z,s}\le\delta_{\sigma_z}/2 \\ 
0, \, {\rm otherwise,}
\end{cases}
\label{eq:Delta}
\end{equation}

\noindent
$\delta_{z,s}=\mid \langle z\rangle-\langle z\rangle_s\mid$, 
$\delta_{\sigma_z,s}=\mid\sigma_z-\sigma_{z,s}\mid$, and $\delta_z$ 
($\delta_{\sigma_z}$) is the bin size in the $\langle z\rangle$ 
($\sigma_z$) axis.

This quantity is displayed in panels (a)-(c) of Fig.\ \ref{fig:Y} 
(normalized to the largest value in each panel) for $M=3$ 
and the different temperatures considered in this work.
These results show that the distribution is peaked at large 
$\langle z\rangle$ values at the lowest temperature and the position 
of the bump moves to lower values as the temperature rises.
The distribution is very broad along the $\sigma_z$ axis, but it 
becomes narrower as the system is heated up.
The position of the peak also moves to lower values as the 
temperature rises from $T=5.5$ MeV to 6.0 MeV.
At $T=5.0$ MeV one observes two bumps along the $\sigma_z$ axis 
which can be explained by the presence of one (two) large fragment(s) 
and two (one) smaller one (ones) in some events, whereas there are 
others in which the three fragments tend to be more similar.
These bumps merge as $T$ increases.
It is important to note that there are no statistical fluctuations 
in these results, since the formulae derived in this work 
are exact within the framework of the model.
Therefore, the patterns observed in the distributions are 
not artificial.

\begin{figure}[tbh]
\includegraphics[width=8.5cm,angle=0]{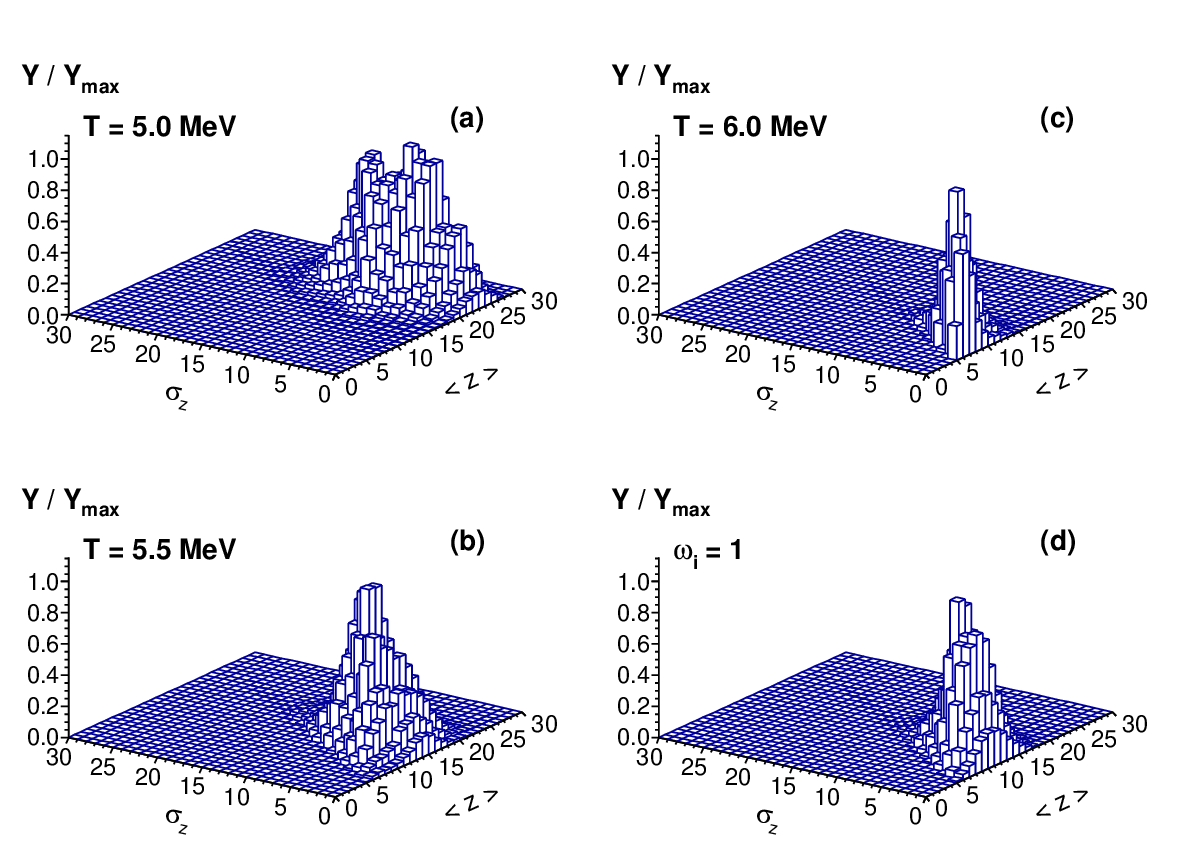}
\caption{\label{fig:Y} (Color online) Panels (a)-(c): Distribution 
of partitions in which the $M=3$ largest fragments (with atomic number 
$z \ge 5$) have average value $\langle z\rangle$ and standard 
deviation $\sigma_z$, for different breakup temperatures.
Panel (d): The partitions are constructed considering only 
combinatorial arrangements of $A_0-Z_0$ neutrons and $Z_0$ protons. 
For details, see the text.}
\end{figure}

Panel (d) of Fig.\ \ref{fig:Y} shows the distribution 
$Y(\langle z\rangle,\sigma_z)$ obtained assuming $\omega_i=1$.
In this case, it is narrower than those obtained at $T\le 5.5$ MeV 
and it is peaked at lower $\langle z\rangle$ and $\sigma_z$ values.
It is more similar to the distribution at $T=6.0$ MeV.
This shows that the statistical weights associated with the 
phase space available to the partitions lead to very important 
deviations from the scenario of fragments populating the partitions 
according to combinatorial arrangements only.

\begin{figure}[tbh]
\includegraphics[width=8.5cm,angle=0]{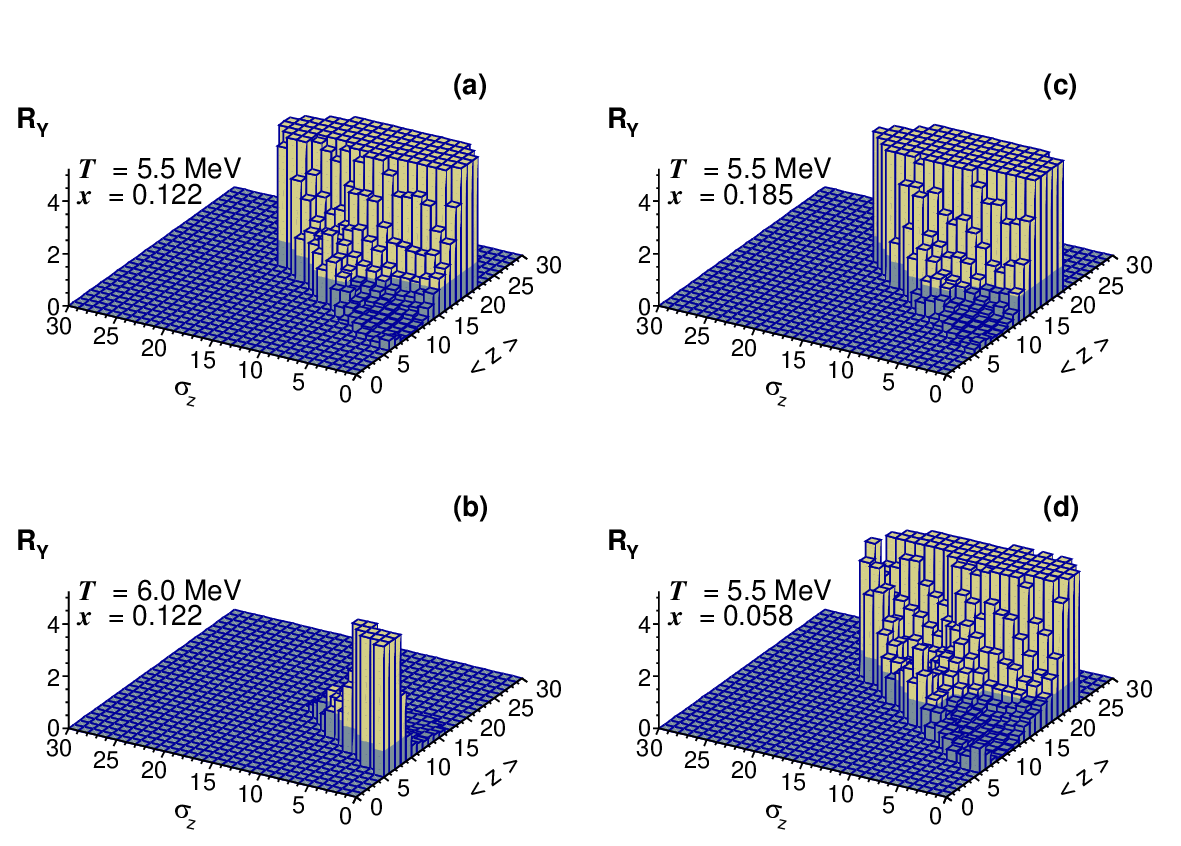}
\caption{\label{fig:R} (Color online) Ratio between the distribution 
$Y(\langle z\rangle,\sigma_z)$ calculated using the full statistical 
weight and the one obtained considering only the 
combinatorial arrangements of the nucleons, at different temperatures 
and asymmetry parameter $x=1-2Z_0/A_0$ of the source of mass number 
$A_0=189$. For details, see the text.}
\end{figure}

For the purpose of providing a more quantitative interpretation 
of this aspect, Fig.\ \ref{fig:R} displays the ratio $R_Y$ between 
$Y(\langle z\rangle,\sigma_z)$ calculated with the full statistical 
weight and the one obtained assuming $\omega_i=1$, 
$Y(\langle z\rangle,\sigma_z,{\omega_i=1})$, in different cases.
In order to eliminate contributions from events that are not representative, we only show the ratios if 
$Y(\langle z\rangle,\sigma_z) \ge 10^{-4}$.
We also limit the vertical scale to $R_Y \le 5$, since very large 
ratios are obtained if $Y(\langle z\rangle,\sigma_z,{\omega_i=1})$ 
is very small.
This is the reason why the distributions look flat in some regions.
Thus, we restrict the analysis to the regions where there is a 
competition between the constraints imposed by the combinatorial 
arrangements of the nucleons and the weights associated with the 
phase space available to the system.

Since at $T=5.0$ MeV, $Y(\langle z\rangle,\sigma_z)$ is 
non-negligible only where $Y(\langle z\rangle,\sigma_z,{\omega_i=1})$ 
is very small, the results at this temperature are not shown in 
Fig.\ \ref{fig:R} and we focus on the highest two temperatures.
In this plot, the blue (dark gray) areas correspond to $R_Y\le 1$ 
whereas the light brown (light gray) ones are associated with 
$R_Y > 1$.
In panels (a) and (b), we exhibit $R_Y$ for $T=5.5$ MeV and 
$T = 6.0$ MeV, respectively.
In the former case, $R_Y < 1$ for $\langle z\rangle\lesssim 15$, 
except for $\sigma_z\gtrsim 10$.
This indicates that the statistical weight of these configurations 
is not large enough to dominate the constraints associated with the 
combinatorial arrangements.
The larger phase space available to configurations which give 
$\langle z\rangle\gtrsim 15$ tips the balance in its favor and 
one observes a rapid rise of $R_Y$ in this region.
The fact that $R_Y > 1$ also for very small values of $\sigma_z$, 
at non-negligible values of $Y(\langle z\rangle,\sigma_z)$, 
reveals that many partition modes with fragments of similar sizes 
contribute to the distribution.
Analogous conclusions also hold for different 
sources' isospin composition as one sees in
panels (c) and (d) which show $R_Y$ for sources, at $T=5.5$ MeV, 
with $A_0=189$ and $x=0.185$ and $x=0.058$, respectively, 
where $x=1-2Z_0/A_0$.
The situation is very different at $T=6.0$ MeV.
The overlap between $Y(\langle z\rangle,\sigma_z)$ and 
$Y(\langle z\rangle,\sigma_z,{\omega_i=1})$ is appreciable and 
$R_Y>>1$ in most of the overlapping region.
This is particularly pronounced at small $\sigma_z$.
Thus, the existence of these very similar fragments is due to statistical considerations rather than to combinatorial 
constraints.
Our results indicate that this dominance of the statistical weights 
over the combinatorial arrangements is sensitive to the excitation 
energy of the source.
Similar conclusions are also obtained in the case of $M=4$.

\begin{figure}[tbh]
\includegraphics[width=8.0cm,angle=0]{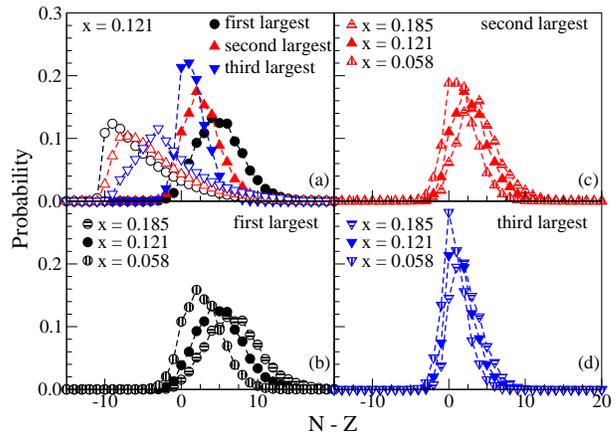} 
\caption{\label{fig:Ynz} (Color online) Distribution of the 
neutron-proton asymmetry of the largest three fragments for 
different values of the sources' asymmetry, at breakup temperature 
$T=5.5$ MeV. 
The open symbols in frame (a) represent the results obtained with 
$\omega_i=1$. For details, see the text.}
\end{figure}

Very different qualitative characteristics of the isospin composition 
of the largest fragments are observed whether one assumes that the 
fragmentation modes are ruled by combinatorial arrangements 
only, or takes into account the full statistical weights.
This is illustrated in panel (a) of Fig.\ \ref{fig:Ynz} which displays 
the $N-Z$ distribution of the $M=3$ largest fragments, produced at 
$T=5.5$ MeV, where $N$ denotes the neutron number.
It reveals that the isospin  properties of the Helmholtz free energy 
leads to neutron richer fragments than considerations based only on 
combinatorial arrangements.
It also shows that, in the former case, the largest fragments tend 
to be more neutron rich than the lighter ones.
This reflects the tendency of nuclei of having equal number of 
neutrons and protons as their sizes diminish.
Panels (b)-(c) of this figure compare the distributions for sources 
of different isospin compositions, for $A_0=189$, at $T=5.5$ MeV.
They show that the sensitivity to the isospin composition of the 
source weakens as the fragments' rank size decreases.

\begin{figure}[tbh]
\includegraphics[width=8.5cm,angle=0]{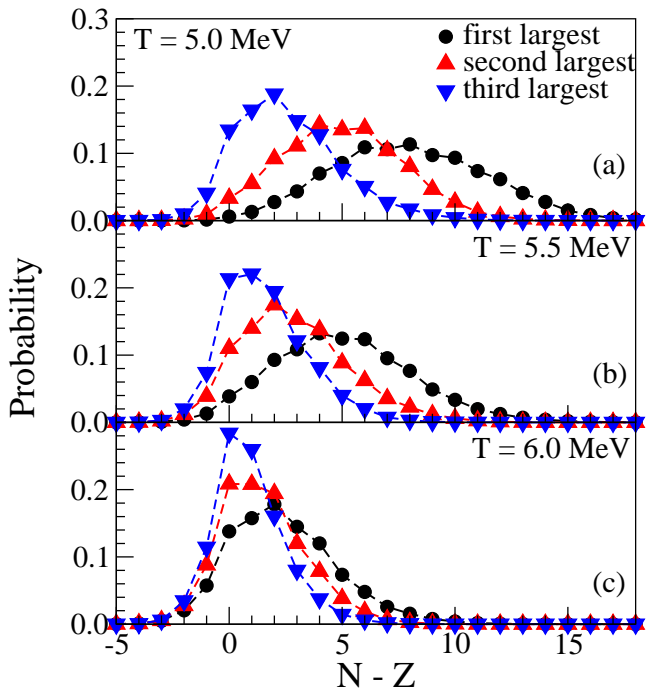}
\caption{\label{fig:YnzT} (Color online) Same as 
Fig.\ \ref{fig:Ynz} for $x=0.121$ and different temperatures. 
For details, see the text.}
\end{figure}

Finally, to investigate the influence of the breakup temperature 
on the $N-Z$ distribution of these largest fragments, this quantity 
is exhibited in panels (a)-(c) of Fig.\ \ref{fig:YnzT} at different 
temperatures.
The results reveal that the distributions become narrower as $T$ 
increases and the peaks move towards $N\approx Z$.
Thus, they may help investigate the asymmetry energy 
term in the equation of state of the source at the 
breakup stage.

\end{section}

\begin{section}{Concluding Remarks}
\label{sect:conclusions}
In the framework of the prompt statistical breakup of a nuclear source 
in thermal equilibrium at temperature $T$, we examined the properties 
of the largest $M$ fragments produced in each fragmentation mode.
We addressed the question of whether the production of 
many similar fragments within an event is ruled by combinatorial 
constraints or by statistical considerations.
To this end, we employed a version of the SMM, presented in 
Refs.\ \cite{ChaseMekjian1995,SubalMekjian}, based on recurrence 
formulae for the statistical weights, and derived expressions which allowed the individual calculation of each partition with $M$ 
fragments of atomic number $z \ge z_{\rm min}$.
Our results suggest that either aspect dominates certain configurations 
and that the balance between them is sensitive to the breakup temperature.
More specifically, larger temperatures lead to larger phase 
space volumes accessible to the system and, therefore, tip 
the balance in favor of the statistical emission and one observes 
many fragments of similar sizes.
However, we found that the partition mode also plays an important role 
as, for a given breakup temperature, the combinatorial arrangements 
of the nucleons dominate in certain configurations, which have access 
to smaller phase space volumes.
We therefore suggest these properties should be further 
experimentally investigated,  and that the configuration 
at the breakup be reconstructed as in 
Refs.\ \cite{freezeOut,primFragsIndra2003,primaryFrags2014,
primaryFrags2014_2,expRecEex2013,fragReconstructionNPA}.
Our results also suggest that the neutron-proton asymmetry of the 
$M$ largest fragments is sensitive to their rank size and that 
this sensitivity weakens as the breakup temperature increases.
The neutron-proton asymmetry of the source is also found to affect 
this property of the $M$ largest fragments.

\end{section}

\begin{acknowledgments}
This work was supported in part by the Brazilian
agencies Conselho Nacional de Desenvolvimento Cient\'\i ­fico
e Tecnol\'ogico (CNPq), by the Funda\c c\~ao Carlos Chagas Filho de
Amparo \`a  Pesquisa do Estado do Rio de Janeiro (FAPERJ),
a BBP grant from the latter. 
We also thank the Uruguayan agencies
Programa de Desarrollo de las Ciencias B\'asicas (PEDECIBA)
and the Agencia Nacional de Investigaci\'on e Innovaci\'on
(ANII) for partial financial support.
This work has been done as a part of the project INCT-FNA,
Proc. No.464898/2014-5.
We also thank the N\'ucleo Avan\c cado de Computa\c c\~ao de 
Alto Desempenho (NACAD), Instituto Alberto Luiz Coimbra de 
P\'os-Gradua\c c\~ao e Pesquisa em Engenharia (COPPE), 
Universidade Federal do Rio de Janeiro (UFRJ), for the use 
of the supercomputer Lobo Carneiro, as well as the Cloud Veneto, 
where the calculations have been carried out.

\end{acknowledgments}

\bibliography{manuscript}
\bibliographystyle{apsrev4-1}

\end{document}